\documentclass[twocolumn, prl]{revtex4-1}
\usepackage{amssymb}

\usepackage{graphicx}
\usepackage{epstopdf}
\usepackage{color}

\usepackage{lineno}
\usepackage{hyperref}
\hypersetup{
	bookmarks=false,         
	unicode=false,          
	pdftoolbar=false,        
	pdfmenubar=false,        
	pdffitwindow=false,     
	pdfstartview={FitH},    
	pdftitle={null},    
	pdfauthor={null},     
	pdfsubject={null},   
	pdfcreator={null},   
	pdfnewwindow=true,      
	colorlinks=true,       
	linkcolor=blue,          
	citecolor=blue,        
	filecolor=blue,      
	urlcolor=blue           
}
\usepackage{multirow}

\begin{document}
	
		\title{Suppression of 1/f noise in solid state quantum devices by surface spin desorption}
		\author{S. E. de Graaf$^{1\dagger}$} \author{L. Faoro$^{2,3}$} \author{J. Burnett$^4$}   \author{A. A. Adamyan$^4$} 
		\author{A. Ya. Tzalenchuk$^{1,5}$} 
		\author{S. E. Kubatkin$^4$} \author{T. Lindstr\"om$^1$}   \author{A. V. Danilov$^4$}
		\affiliation{$^1$National physical laboratory, Hampton Road, Teddington, TW11 0LW, UK}
		\affiliation{$^2$Laboratoire de Physique Theorique et Hautes Energies, CNRS UMR 7589, Universites Paris 6 et 7, place Jussieu 75252 Paris, Cedex 05 France}
		\affiliation{$^3$L. D. Landau Institute for Theoretical Physics, Chernogolovka, 142432, Moscow region, Russia}
		\affiliation{$^4$Department of Microtechnology and Nanoscience, Chalmers University of Technology, SE-412 96 G\"oteborg, Sweden}
		\affiliation{$^5$Royal Holloway, University of London, Egham, TW20 0EX, UK}
		\affiliation{$^\dagger$\rm sdg@npl.co.uk}
		
\begin{abstract}

Noise and decoherence due to spurious two-level systems (TLS) located at material interfaces is a long-standing issue  in solid state quantum technologies. Efforts to mitigate the effects of TLS have been hampered by a lack of surface analysis tools sensitive enough to identify their chemical and physical nature. 
Here we measure the dielectric loss, frequency noise and electron spin resonance (ESR) spectrum in superconducting resonators and demonstrate that desorption of surface spins is accompanied by an almost tenfold reduction in the frequency noise. 
We provide experimental evidence that simultaneously reveals the chemical signatures of adsorbed magnetic moments and  demonstrates their coupling via the  electric-field degree of freedom to the resonator, causing dielectric (charge) noise in solid state quantum devices. 
		
\end{abstract}

\maketitle

As the complexity of solid state quantum circuits continues to increase, so do the challenges  to both fabrication technology and materials science\cite{devoret2013}.
Improved device and systems engineering has lead to  material imperfections  being a dominant source of noise and decoherence, and further improvements in material properties and a better understanding of the underlying materials physics are needed to make technologies such as large scale solid state quantum computing feasible\cite{paladino2014, devoret2013, muller2017}.
The enhanced sensitivity of superconducting qubits and resonators has revealed that materials once considered to be near-perfect crystals, actually contain sufficient imperfections to behave as disordered systems.   
One unexpected consequence of the enhanced sensitivity to disorder of quantum devices was their ability to verify detailed predictions of the  Standard Tunnelling Model (STM)\cite{anderson, phillips}. 
The STM, originally developed to model the low-temperature acoustical and electromagnetic properties of glasses, assumes the presence of a large ensemble of two-level systems (TLS) which can absorb energy via their electric dipole moments, leading to dissipation via subsequent phonon decay. 
 TLS  affect the performance of a many different solid state devices including superconducting resonators and qubits\cite{muller2017}, field-effect transistors\cite{pearl2016}, single charge devices\cite{arsalan2014} and ion traps\cite{iontrap}. Understanding and removing TLS is therefore important for a wide range of applications in solid state physics, materials science and chemistry. 

While the origin of these TLS remains elusive, engineering advances have reduced TLS loss to a level where  most remaining  TLS are  located at or in thin surface oxide layers\cite{steffen2017, chu2016, kumar2016, quintana2017, burnett2014, neill2013}. In this regime the STM fails\cite{burin2015,faoro2015}. Remarkably, measurements of TLS-induced 1/f noise at low temperatures show an increasing noise  $\propto T^{-(1+\mu)}$, with $\mu\sim 0.3$  found in both resonators\cite{burnett2014, burnett2016, skacel2014} and qubits\cite{lisenfeld2010}. This dependence, clearly different from the vanishing $T^3$ dependence of the STM, is a signature of strong long-range TLS interactions.
Furthermore, high quality (high-Q) resonators typically show a much weaker power dependence of the quality factor than what is predicted by the STM\cite{burnett2016, faoro2012, khalil2010, lindstrom2009}.  
This prompted the development of a Generalised Tunnelling Model (GTM)\cite{faoro2015} which takes into account strong dipole-dipole \textit{interactions} between TLS\cite{lisenfeld2015}, successfully capturing the observed physics. 

\begin{figure*}
	\includegraphics[scale=1.28]{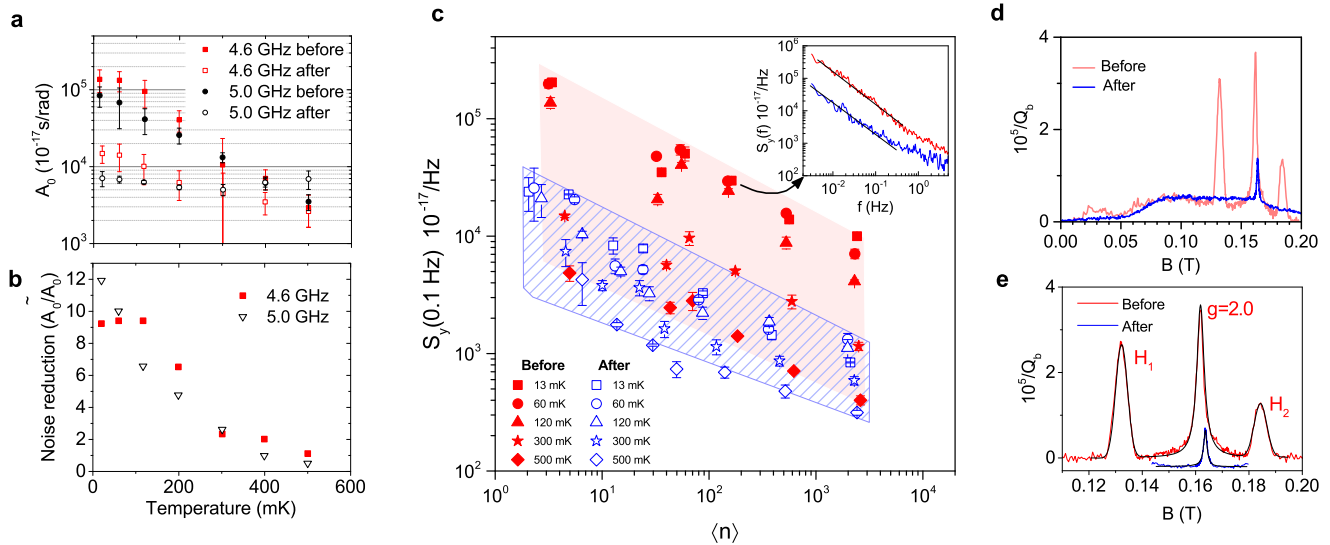}	\vspace{-12mm} 	
	\caption{Reduction of noise due to surface spin desorption.  a) The extracted noise amplitude $A_0$ in the low power limit, obtained from the frequency noise spectral density  $S_y(\langle n\rangle, T,f) = A_0(\langle n\rangle, T)/2\pi f$ as a function of temperature in two resonators before and after spin desorption. b) The change in noise amplitude before/after $=A_0/\tilde{A}_0$ vs temperature. a) and b) are extracted from the full power and temperature dependence of the measured frequency noise power spectral density in c).   c) Frequency noise power spectral density $S_y(f) = S_{\delta \nu}(f)/\nu_0^2$ at $f=0.1$ Hz for the $\nu_0=4.6$ GHz resonator (see Supplemental for 5.0 GHz resonator data).  Red solid markers are before, and blue hollow markers are after spin desorption respectively. Shaded regions are a guide for the eye. The inset shows a typical 1/f noise spectral density at 60 mK before ($\langle n\rangle\sim 200$) and after ($\langle n\rangle\sim 100$). Straight black lines are $1/f$.
	d) The full ESR spectrum measured at 10 mK before and after for the 4.6 GHz resonator, verifying that a large number of spins have been removed and e) shows the same data zoomed in together with fit to theory (black lines) and the two hydrogen hyperfine peaks ($H_1$ and $H_2$) indicated together with the free electron peak $g=2.0$ (see Supplemental material). Wide background has been subtracted and curves have been offset for clarity. }
	\label{fig2}
\end{figure*}

Despite this success existing models do not give information about the chemical nature of surface TLS; something that is clearly needed for their  mitigation. Directly studying the chemical nature of TLS using established surface analysis techniques remains extremely challenging. One reason is that the density of TLS is very small, $<$1\% of surface sites, and likely comprised of very light elements\cite{holder2013}, weakly adsorbed molecules\cite{lee2014, wang2015, lordi2017} or electronic defect states\cite{desousa2007}. These are easily introduced by exposing devices to ambient conditions\cite{kumar2016, degraaf2017}, inhibiting the use of many surface analysis techniques\cite{kelber2007}. 
In constrast to charge TLS, magnetic dipoles as sources of flux noise  originate from a bath of paramagnetic surface spins\cite{sendelbach2008, quintana2017, kumar2016}, and can therefore be identified by electron spin resonance (ESR) techniques using  sensitive tools derived from solid state quantum technologies\cite{degraaf2017, quintana2017, samkharadze2016}. Identifying  TLS that  couple through their charge degree of freedom is much more challenging due to the lack of direct identification methods that can reveal chemical fingerprints.

In this work we show that changes observed in noise and loss measurements of superconducting resonators  directly correlate with ESR data, which reveals important new clues about the chemical and physical nature of surface TLS. 
We further show that desorbing spins with a simple annealing treatment leads to a reduction of the frequency noise by almost an order of magnitude (see Figure \ref{fig2}a and \ref{fig2}b).  This also allows us to directly identify the origin of the TLS responsible for noise as atomic scale electric dipoles; some of which are comprised of physisorbed atomic hydrogen\cite{degraaf2017}, while others are associated with free radicals.
Our results suggest that these paramagnetic species not only cause a fluctuating magnetic environment\cite{quintana2017}, but also are responsible for dielectric (charge) noise.

\section{Experiments}
We simultaneously measure the $1/f$ frequency noise and dielectric losses as a function of temperature and driving power (average photon number $\langle n\rangle$)  of two NbN superconducting resonators (with frequencies ${\nu_0=4.6}$ GHz and $5.0$ GHz)\cite{aplpaper} patterned on the same {\it c}-cut Al$_2$O$_3$ substrate. 

The full high sensitivity ESR spectrum is subsequently obtained at $T=10$ mK 
by measuring the quality factor of the resonator as a function of applied magnetic field and the zero field loss is subtracted to obtain the magnetic field induced loss $Q_b^{-1}$\cite{jap2012}.
 We then anneal the device at moderate temperature (300$^\circ$C), a technique that has shown to remove some of the spins native to the surface of the device\cite{degraaf2017}. The same noise and loss measurement protocol is repeated in a second measurement and finally the ESR spectrum is measured again, confirming the successful removal of some of the spins.
Throughout this paper we refer to these two consecutive measurements as 'before' and 'after' spin desorption respectively.

\begin{figure*}[!ht]
	\includegraphics[scale=1.12]{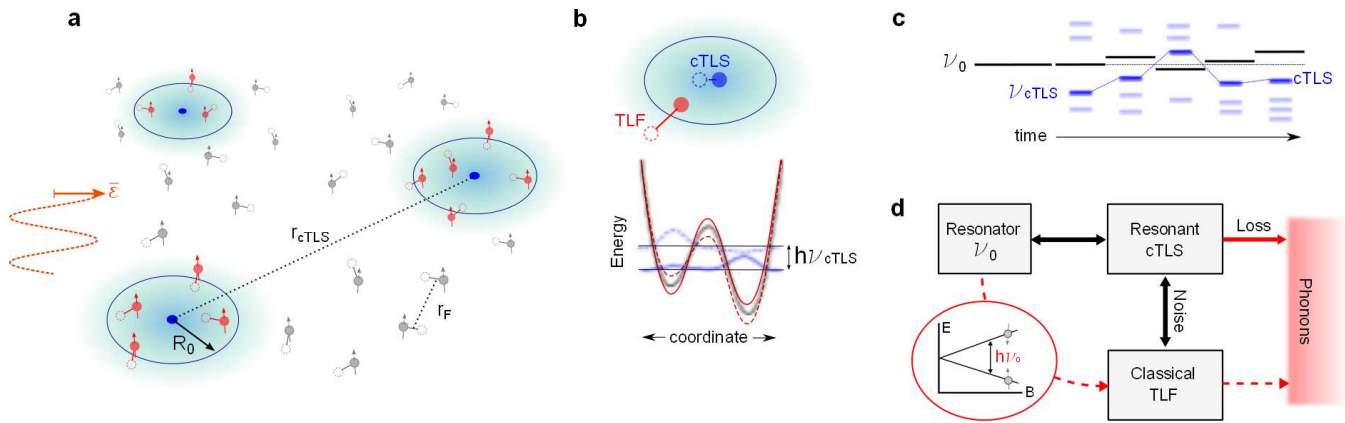} 
	\caption{Mechanism of dielectric (frequency) noise and loss in high-Q superconducting resonators. a) A smaller number of coherent cTLS (blue) on average separated by a distance $r_{\rm cTLS}\sim 1$ $\mu$m couple to the oscillating electric field component $\vec{\mathcal{E}}$ of the resonator. Classical thermally activated TLFs (red)  in $R_0\sim 80 $ nm proximity of the cTLS generate noise while other thermally activated TLFs (grey) contribute to the cTLS line-width and the total density of TLS detected in ESR measurements. Typical distances between thermally activated TLF (at $T\sim60$ mK) are $r_F\sim 100$ nm. b)  TLF inside the interaction volume of the cTLS modify the tunneling potential of the cTLS, resulting in the cTLS energy drift. c) The resonantly coupled cTLS have energy level splittings near the resonance frequency $\nu_0$. This splitting fluctuates in time, perturbing the resonator frequency via its coupling to the electric dipole associated with the cTLS. d) The conceptual representation of the GTM where noise and loss channels are indicated (see text). The ESR measurement enables identification of TLFs via the new dissipation channel, indicated by dashed lines, arising when the spins are in resonance with the microwave field.} \label{fig:sketch}
\end{figure*}

The frequency noise is measured in two resonators  using a high precision dual Pound locking technique adapted from frequency metrology\cite{tobiaspound} that continuously monitors the centre frequency of the resonators.
Values for the dielectric loss tangent $\tan\delta_0$ before and after annealing are extracted from quality factor measurements at low power, and from an independent measurement of the temperature-dependent frequency shift of the resonators we find the intrinsic loss tangent $\tan\delta_i$. For further details see Methods.

\section{Results}
The main result of this work is shown in Figure \ref{fig2}. In summary, after annealing and  desorption of  surface spins we observe almost an order of magnitude reduction (on average 9.1 and 8.4 times for the two resonators respectively) in the frequency noise spectral density (Figure \ref{fig2}a-c) for both measured resonators at the lowest temperatures. 

The reduction in noise is observed together with a reduction in number of surface spins.
Figures \ref{fig2}d and \ref{fig2}e display the ESR spectrum measured in-situ after collecting all the noise data, before and after annealing.  The measured ESR spectrum reveals the presence of atomic hydrogen on the Al$_2$O$_3$ surface originating from water dissociation\cite{hass1998} and electronic charge states (with a g-factor of 2.0), likely due to absorption of oxygen radicals on the surface in accordance with previous findings\cite{degraaf2017, kumar2016}. 
An initial density of $n_H = 2\cdot 10^{17} $ m$^{-2}$ hydrogen spins is completely removed and we extract a reduction in spin density due to the central peak from $n_e = 0.91 \cdot 10^{17}$ m$^{-2}$ to $\tilde{n}_e = 0.17 \cdot 10^{17}$ m$^{-2}$ spins/m$^2$, a factor of 5.3. The wide background plateau remained unchanged. 

 Intriguingly, in contrast to the tenfold reduction in noise, we find that the intrinsic loss tangent $\tan \delta_i$ is only reduced by 30\% after surface spin desorption. 
For each resonator we also measured the power and temperature dependence of the quality factor, from which we also see only a very small reduction in the loss (see Table \ref{tab1} for exact values).

\section{Discussion}
This small reduction in loss but large reduction in noise can be explained within the framework of  strongly interacting TLS and the GTM, which naturally partitions the TLS as two distinct entities, one predominantly responsible for loss and one for noise. 
The microscopic picture is the following. Associated with each TLS there is a fluctuating dipole  $d_0$ that couples to the applied microwave electric field ${\cal E}$ from the resonator. 
Among the TLS we can distinguish between coherent (quantum)  electrical dipoles (cTLS) that are characterized by fast transitions between their states and relatively small decoherence rates, and slow classical fluctuators (from now on referred to as TLF) that are characterized by decoherence times shorter than the typical time between the transitions. 
The picture is sketched in Figure \ref{fig:sketch}a with typical distances between thermally activated (excited) cTLS and TLF as inferred from our measurements.

At low temperatures, slow fluctuators weakly coupled to cTLS mainly contribute to the dephasing of the high energy  cTLS and are responsible for their line-width $\Gamma_2$\cite{faoro2015, lisenfeld2016}. Slow fluctuators that are located close to the cTLS, and therefore are strongly coupled, shift the cTLS energy by an amount larger than $\Gamma_2$. These fluctuators create highly non-Gaussian noise that cannot be regarded as a contribution to the line-width.  For resonant cTLS, having an energy splitting ${E \approx \hbar \nu_0}$, the interaction with a few strongly coupled TLF translates to the energy of the cTLS drifting in time, as illustrated in Figure \ref{fig:sketch}b and c; it is this drift that ultimately generates $1/f$ noise in the resonator\cite{faoro2015}.

The intrinsic loss  (at low fields) on the other hand arises from direct phonon relaxation from the resonantly coupled cTLS, and depends only on the number of cTLS, as shown in Figure \ref{fig:sketch}d. 
Within the framework of the GTM our experimental findings of a small reduction in loss and a dramatic reduction in noise imply that desorption of surface spins did not affect the density of cTLS, instead the surface spins can be attributed to the TLF.

\begin{figure}[!t]
	\includegraphics[scale=0.6]{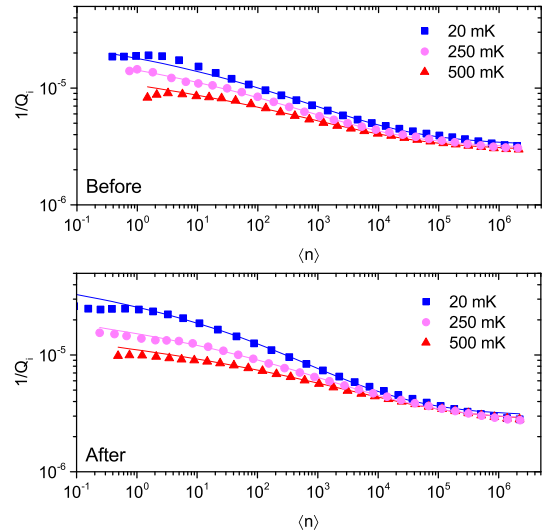}	
	\caption{Resonator quality factor. Inverse internal quality factor as a function of number of photons in the 5 GHz resonator. Solid lines are fits to the logarithmic power dependence of equation (\ref{eq:logpower})  for $\langle n\rangle \gtrsim 50$. Extracted values are reported in Table \ref{tab1}.} \label{fig:qilog}
\end{figure}

The conceptual picture of these two separate TLS communities is further supported by additional experimental findings:
for a homogeneous bath of non-interacting TLS (STM)
 we expect $Q_i(\langle n\rangle)\sim \langle n \rangle^\alpha$ with $\alpha = 0.5$.
 The observed dependence is much weaker: a fit to a power law returns $\alpha\approx 0.2$ for both resonators before and after desorption (see Supplementary). On the other hand, for interacting TLS we do expect a weak logarithmic dependence of the microwave absorption on stored energy in the resonator\cite{faoro2012}
\begin{equation} 
\frac{1}{Q_i(\langle n\rangle)} = P_\gamma F \tan\delta_i \ln\left(C \sqrt{\frac{|n_c|}{|\langle n\rangle|}}+c_0 \right).\label{eq:logpower}
\end{equation}
Here $C$ is a constant, $c_0$ accounts for power-independent losses, F is a geometric filling factor and  $\displaystyle{P_\gamma}$ is a normalization factor that depends on the spectral density of TLF switching rates.
In Figure \ref{fig:qilog} we show that our data fits very well to this logarithmic power dependence.
Interestingly, we find that $P_\gamma$ increases  after spins were removed. This implies that the remaining slow fluctuators have a narrower range of switching rates and are likely different in nature than the spins that were desorbed.

Independently, another important indication of the applicability of our model is given by the analysis of the temperature dependence of the  $1/f$ noise spectrum. The interaction gives a vanishing density of states for cTLS at low energies, $P(E) \propto E^\mu$ with $0<\mu<1$, and this results in a scaling of the noise spectrum with temperature $S_y(T) \propto T^{-(1+ 2 \mu)}$ (for ${T < h\nu_0/k_B}$). In agreement with previous studies\cite{burnett2016, burnett2014, skacel2014, lisenfeld2010} we find  ${\mu \approx 0.3}$ (see Supplemental material), both before and after spin desorption. This is further evidence that desorption only affects the number of slow TLF present on the sample.

We now combine all available data to produce a qualitative picture (as sketched in Figure \ref{fig:sketch}) of the microscopic properties of the cTLS and TLF, by taking the GTM beyond the original assumptions of identical densities and dipole moments of cTLS and TLF\cite{faoro2015, burnett2016}. The details of this theory and analysis can be found in the Supplemental material, here we only summarize the results. 

Assuming the dipole moment for resonant cTLS to be on the atomic scale,  ${d_0 = 1 \text{ e\AA} \sim 5 \text{D}}$ (i.e. similar to what was previously deduced from spectroscopy measurements\cite{sarabi2016, martinis2004}), we arrive at dipole-dipole interaction strength ${U_0 \approx 15 \text{ K}\text{nm}^3}$.  
Before spin desorption, we find from the intrinsic loss tangent the cTLS line-width ${\Gamma_2 \sim 20 \text{ MHz}}$ at ${T=60 \text{ mK}}$ (see Supplemental material), which translates into the density of resonant cTLS ${\rho_{TLS} \approx 15 \text{ GHz}^{-1} \mu\text{m}^{-2}}$ in agreement with Ref. \cite{lisenfeld2016}, where the authors found $\sim 50$ resonant cTLS per $\mu\text{m}^{2}$ in the frequency range ${3-6}$ GHz, i.e. resonant cTLS are  located at a typical distance $r_{cTLS} \sim 1$ $\mu\text{m}$ from each other, similar to the densities found in qubit tunnel junctions\cite{martinis2004}. 

Next, the measured amplitude of the noise $A_0$ can be related to the density of thermally activated (fluctuating) TLF and their dipole moment $d_F$. We find $\displaystyle{\frac{d_F}{d_0} \rho_F \approx 5\pm4 \cdot 10^{-3} \text{ nm}^{-2}}$. 
The thermally activated TLF constitutes a fraction $T/W$ of the total number of TLF, where $W$ is the bandwidth of the distribution of TLF energy level splittings. For weakly absorbed spins it is reasonable to expect that ${W\sim 100}$ K, limited by the observed desorption energy. From the total spin density measured by ESR we have ${n_e+n_H \approx 3 \cdot 10^{-1} \text{ nm}^{-2}}$. 

Combining these estimates, assuming all the TLFs are the observed spins, we have for the density of thermally activated TLF ${\rho_F = (n_e+n_H)(T/W)\sim 2 \cdot 10^{-4} \text{ nm}^{-2}}$ (i.e. thermally activated TLFs are separated by an average distance ${r_F \sim 100 \text{ nm}}$) and ${d_F/d_0 \sim 30\pm25}$. The large uncertainty in $d_F/d_0$  stems from its strong dependence on the filling factor ($\propto F^3$) and the volume where the TLF are situated, which both cannot be accurately estimated.
However, the message of this order of magnitude estimation is that the assumption that all TLS are the observed spins is indeed plausible. Furthermore, the dipole moment of a surface TLF is likely larger compared to that of TLS in the bulk, as would be expected since the physisorbed and easily desorbed spins are likely to move larger distances.

After spin desorption the noise amplitude decreases by a factor $\sim 10$, the loss is only reduced by ${\sim 30 \%}$ and the normalization constant $P_\gamma$ increases $\sim 65\%$ due to lower TLF switching rates. From this we can finally find a corresponding change in the density of TLF before and after spin desorption
\begin{equation}
\frac{\rho_F(T)}{\tilde{\rho}_F(T)}= \frac{A_0\tilde{P}_\gamma\tan\delta_i}{\tilde{A}_0P_\gamma\tan\tilde{\delta_i}}   =  15.23 \quad \text{and} \quad  17.7,
\end{equation}
for the two resonators. Here we denote quantities for the 'after' measurement by the tilde symbol. These values correlate remarkably well with the change in the total number of spins in the three ESR  peaks $ (n_e+n_H)/\tilde{n}_e=  17.1$ (4.6 GHz resonator), and again indicates that spins contribute to the frequency noise in our high-Q superconducting resonators and take on roles as slow (mobile\cite{hass1998}) fluctuators.

\section{Conclusions}
Based on the experimental evidence from the loss, noise and ESR spectrum, all obtained on the same device, we have found that surface spins that are known to give rise to magnetic noise in quantum circuits\cite{sendelbach2008, quintana2017, degraaf2017}  are also responsible for the low frequency dielectric noise of the resonator. These spins, remarkably present in densities also inferred to be responsible for flux noise in SQUIDs and qubits\cite{sendelbach2008},    take on roles as slow classical fluctuators that cause an energy drift of resonant coherent TLS. Removing a majority of these spins gives an almost tenfold reduction in dielectric noise.

In our device the observed surface spins constitute  weakly physisorbed atomic hydrogen together with free radicals ($g=2$). We note that the nature of the $g=2$ spins is still not entirely clear. A large portion can be associated with surface adsorbates, likely oxygen radicals\cite{degraaf2017,kumar2016}, or other light molecular adsorbents\cite{lee2014, lordi2017}. The remaining fraction of free radicals may be a result of insufficient annealing or they may be of a different chemical or physical origin with much higher desorption barriers. Another possibility is that the remaining more robust localised charges and cTLS are intrinsic to the Al$_2$O$_3$ surface itself\cite{desousa2007, hass1998}, more resembling "bulk" defects\cite{bedilo2014}.
Nevertheless, our approach reveals a new aspect of the noise in solid state quantum devices as we show that observed magnetic dipoles, with their fingerprint revealed through state-of-the-art surface analysis using in-situ micro-ESR, couple via the electric field degree of freedom and give rise to dielectric noise.

Similar physics is expected for a wide range of oxide surfaces relevant for quantum technologies.
The importance of magnetic moments has previously been widely overlooked in resonators since  electrical dipoles have been considered the dominating mechanism for dielectric noise. Our results instead indicate that while having a small influence on power loss, these spins (and their associated electric dipoles) constitute a major source of noise and dephasing in modern high coherence solid state devices by their proximity to coherently coupled resonant cTLS, and our results hint at a connection between the similar densities found for sources of flux\cite{sendelbach2008}, charge\cite{astafiev2004, arsalan2014} and dielectric noise in quantum circuits.

\begin{table}
	\begin{tabular}{r| c | c| c | r }
		Quantity & Unit & Before & After & Note\\
		\hline\hline
		
		\multirow{ 2}{*}{Spin density$^\dagger$} &  \multirow{ 2}{*}{$ 10^{17}$m$^{-2}$} & $0.91$ & $0.17$ & $g=2$\\
		& & $2.0$ & 0 & H\\\hline

		\multirow{ 2}{*}{$F\tan\delta_i$} & \multirow{ 2}{*}{$\times 10^{-6}$} & $10.6\pm 0.15$ & $7.44\pm 0.13$ &4.6 GHz\\
		& & $10.4\pm 0.27$& $7.69\pm 0.12$ &5.0 GHz\\\hline

		\multirow{ 2}{*}{$P_\gamma F\tan\delta_i$} & \multirow{ 2}{*}{$\times 10^{-6}$} & $4.2\pm0.24$ & $4.9\pm 0.1$ & 4.6 GHz\\
		& & $5.4\pm 0.6$ & $6.5\pm0.6$ & 5.0 GHz\\\hline
		\multirow{ 2}{*}{$P_\gamma$} & \multirow{ 2}{*}{} & $0.39\pm 0.02$ & $0.66\pm 0.02$ & 4.6 GHz\\
		& & $0.52\pm 0.06$ & $0.84\pm 0.08$ & 5.0 GHz\\\hline

		\multirow{ 2}{*}{$\alpha$} &  & $0.20\pm0.024$ & $0.18\pm 0.037$ & 4.6 GHz\\
		& & $0.27\pm 0.02$ & $0.22\pm0.038$ & 5.0 GHz\\\hline

		$2 \mu$ & & $0.64\pm 0.50$ & $0.43\pm 0.21$ & 4.6 GHz\\\hline
		
		\multirow{ 2}{*}{$A_0/2\pi $} & \multirow{ 2}{*}{$10^{-17}$} & $2.2\pm0.3\cdot 10^4$ & $2.4\pm0.4\cdot 10^3$ & 4.6 GHz\\
		& & $1.2\pm0.4\cdot 10^4$ & $1.1\pm0.3\cdot 10^3$ & 5.0 GHz\\\hline\hline

	\end{tabular}
	\caption{Extracted parameters from ESR and noise/loss measurements. For a detailed description of each parameter see Refs. \cite{degraaf2017, burnett2016, faoro2015} and the  supplemental material. $^\dagger$For the 4.6 GHz resonator. Where indicated, deviations are 95\% confidence bounds or propagated errors thereof from fitting.}
	\label{tab1}
\end{table}

\section{Acknowledgements}
This work was supported by the UK government's Department for Business, Energy and Industrial Strategy. The authors would like to thank S. Lara-Avila for assistance with fabrication. LF acknowledges support by ARO grant W911NF-13-1-0431 and by the Russian Science Foundation grant  \#14-42-00044. AD acknowledges support from VR grant 2016-04828.

\section{Author contributions}
SdG, SK, AD, TL and AT conceived the experiment. AA prepared and treated the samples. SdG performed the noise and ESR measurements with assistance from TL. SdG analysed the data and LF articulated the theory, both with inputs from JB, AT and TL. All authors discussed the results. SdG, LF and TL wrote the manuscript with input from all authors. 


\section{Methods}
{\it Sample preparation. }
Sapphire substrates were annealed in situ at high temperature, $800^\circ$C, for  20 minutes prior to deposition of 2 nm NbN. After cooling down to $20^\circ$C, an additional 140 nm NbN was sputtered. 
Resonators were patterned using electron beam lithography (UV60 resist, MF-CD-26 developer, DI water rinse) and subsequent reactive ion etching in a NF$_3$ plasma. Resist was removed in 1165 remover followed by oxygen plasma treatment. Resonator designs were identical to those reported in Ref. \cite{aplpaper}.
After the first round of noise measurement the same sample was warmed up, shipped from UK to Sweden, and heated in vacuum to $\sim 300^\circ$C for 15 minutes to desorb surface spins, then shipped back to the UK, and mounted in the same cryostat with the same noise measurement setup $\sim 72$ hours later. Remarkably, the detrimental surface spins are not re-introduced even after this time.

{\it Measurement setup. }
We used a cryogen-free dilution refrigerator with a base temperature of 10 mK and a 3-axis superconducting vector magnet for noise and ESR measurements. The cryostat was equipped with heavily attenuated coaxial lines, cryogenic isolators and a low noise high electron mobility transistor (HEMT) amplifier with a noise temperature of $\sim 4$ K. All noise measurements were performed with the leads to the vector magnet completely disconnected. Only after completion of noise measurements the magnet was connected to measure the ESR spectrum. The plane of the superconductor thin-film was found to high precision ($<0.1^\circ$) by applying a small field and carefully tilting the angle of the applied field while finding the maximum of the resonance frequency of the resonators. 
ESR measurements were performed by sweeping the magnetic field and measuring the characteristics of the resonators using a vector network analyser.
Noise measurements were performed using a Pound locking technique\cite{tobiaspound} that tracks the resonance frequency (and its fluctuations) in real-time. For a detailed explanation of the technique, see supplemental material.


\clearpage\appendix

\section*{Supplemental material}
\renewcommand\thefigure{\thesection S\arabic{figure}}    
\setcounter{figure}{0} 
\renewcommand\theequation{S\arabic{equation}}
\subsection{Model for interacting TLS}

Our experiments on noise and loss indicate that interactions between TLS are important. They also demonstrate that while the spin desorption procedure significantly affects the magnitude of the noise it has only a minor effect on the intrinsic loss tangent. In this section we discuss the full microscopic model of interacting TLS and their physical origin that follows from the data. 

The fact that surface spin removal has a small effect on the loss tangent implies that the spins do not contribute significantly to the loss at high frequency and thus are not a part of the resonant cTLS ensemble. However, the reduced noise implies that the spins are a significant host of the bath of slow fluctuating dipoles (TLF). The ESR spectrum also indicates that the desorbed spins (both H and free electron states) can be highly mobile  and can tunnel a long distance, i.e. they can easily serve as the dominant fraction of slow fluctuating dipoles.

Coherent cTLS are associated with fluctuating dipoles $d_0$ and are described by the Hamiltonian $H_{TLS}=\frac{\Delta}{2} \sigma^z+\frac{\Delta_0}{2} \sigma^x$ characterized by an asymmetry $\Delta$, tunneling matrix element $\displaystyle{\Delta_0}$, and ${\sigma^a, a=x,y,z}$ are the Pauli matrices. In the rotated basis, the Hamiltonian is simply $H_{TLS}=E S^z$, where ${E=\sqrt{\Delta^2+\Delta_0^2}}$ is the TLS energy splitting and $\displaystyle{S^z=\frac{1}{2} (\cos \theta \sigma^z+\sin \theta \sigma^x)}$ with ${\tan\theta =\Delta_0/\Delta}$. The interaction strength is set by the dipole-dipole interaction scale ${U_0=d_0^2/\varepsilon_h}$, where $\varepsilon_h=10$ is the dielectric constant of the host medium. As a consequence of this interaction, the TLS density at low energies is ${P_{TLS}(E,\sin\theta) = \frac{P_0}{\cos\theta \sin\theta} \left (\frac{E}{E_{\rm max}}\right )^\mu }$, where ${\mu<1}$ is a small positive parameter.  Among coherent TLS we distinguish high, ${E \gg k_BT}$, and low ${E\leq k_BT}$ (thermally activated) energy TLS. In addition, some TLS can be (near) resonant with the resonator, $E\sim h\nu_0$.

The slow fluctuators are represented by classical fluctuating dipoles with moment $d_F$, characterized by switching rates $\gamma$ with a  probability distribution ${P_F(E,\gamma)=P_0^F/\gamma}$ and ${\gamma_{min} \ll \gamma \ll \gamma_{max}}$. Such a distribution for the switching rates appears naturally for thermally activated tunneling. 

The loss in a high quality resonator is caused by fluctuating dipoles with energies close to the resonator frequency $\nu_0$. In the regime of low temperature, ${k_BT \ll h\nu_0}$, the resonant cTLS have a small dephasing width due to their interaction with thermally activated TLS and TLF. This width is given by\cite{supp_faoro2015}
\begin{equation}
\Gamma_2 = \ln \left ( \frac{\Gamma_1^{max}}{\Gamma_1^{min}} \right ) \chi \frac{T^{1+\mu}}{\nu_0^\mu},
\label{dephrate}
\end{equation}
where ${\chi=P_0  U_0 \left (\frac{\nu_0}{E_{max}}\right )^\mu}\approx \tan\delta_i$ 
is a dimensionless parameter, obtained directly from loss tangent measurements, that controls the effect of the interaction on the resonant cTLS. $\Gamma_1^{max}$ and $\Gamma_1^{min}$ are  the minimum and maximum relaxation rates of these cTLS respectively. Direct measurements give ${\Gamma_1^{max} \approx 10^4 \text{ s}^{-1}}$ for the thermally activated cTLS at ${T \approx 35 \text{ mK}}$\cite{supp_lisenfeld2016}. The precise value of ${\Gamma_1^{min}}$ for thermally activated cTLS is not known. However, the electrical noise data shows that $1/f$ noise generated by these cTLS extends to very low frequencies ${f  \leq 1 \text{ mHz}}$ beyond which the dependence changes. This implies that ${\Gamma_1^{min} \approx 10^{-3} \text{ s}^{-1}}$, such that ${\ln \left ( \Gamma_1^{max}/\Gamma_1^{min} \right ) \approx 20}$. 

The total number of resonant cTLS in a volume $V_h$ ($=2.4$ and 2.2 $\cdot 10^{-16}$ m$^3$ for the two resonators respectively) of host material can then be estimated from the measured loss tangent as ${{\cal N}_{res} = \frac{\chi}{U_0} V_h \Gamma_2}$; their average distance is ${r_{cTLS} \sim  \left (\chi \Gamma_2/U_0 \right )^{-1/3}}$ in bulk material and  ${r_{cTLS} \sim  \left (d \chi \Gamma_2/U_0 \right )^{-1/2}}$ in a thin film of thickness $d \ll r_{cTLS}$. 

The noise in the resonator is due to the slow TLF that interact strongly with these resonant cTLS and create highly non-Gaussian noise, that cannot be regarded as a contribution to $\Gamma_2$. These TLF are located at distance $r< R_0$, where ${R_0^3= \frac{d_F}{d_0} \frac{U_0}{\Gamma_2}}$.  Their switchings bring the cTLS in and out of resonance with the resonator leading to $1/f$ frequency noise. The number of thermally activated TLF strongly coupled to a resonant cTLS is ${{\cal N}_F(T)=P_0^F \frac{4 \pi}{3} R_0^3 T}$ and their average distance is ${r_F \sim (P_0^F T)^{-1/3}}$. If the total number of such fluctuators, ${{\cal N}_F^{tot}(T) =  {\cal N}_{res} {\cal N}_F(T) \gg 1}$, the frequency noise spectrum of the resonator can be expressed as a superposition of Lorenztians generated by the switching of the TLF strongly coupled to the resonant cTLS. 

In the limit of weak electric field $\vec{{\cal E}}$  we find that the noise is given by\cite{supp_faoro2015}
\begin{equation}
\frac{S_{\delta \nu}}{\nu_0^2}= \frac{8}{15} \langle d_0^4 \rangle  \frac{\chi}{U_0 \Gamma_2} {\cal F}(\vec{{\cal E}}) {\cal N}_F(T)  \int_{\gamma_{min}}^{\gamma_{max}} \frac{\gamma P(\gamma)}{\gamma^2+\omega^2} d\gamma . 
\label{nois}
\end{equation}
Here ${P(\gamma)=P_\gamma/\gamma}$ is the normalized distribution function of slow fluctuators with $\displaystyle{P_\gamma=\ln^{-1}[\gamma_{max}/\gamma_{min}]}$  and 
$\displaystyle{{\cal F}(\vec{{\cal E}})=\frac{\int_{V_h} |\vec{{\cal E}}|^4 dV}{4 (\int_{V} \varepsilon_0 |\vec{{\cal E}}|^2 dV)^2} \approx \frac{F^2}{ \varepsilon_h^2 V_h}}$ where we introduced the filling factor $\displaystyle{F= \frac{\int_{V_h} \varepsilon_h  |\vec{{\cal E}}|^2 dV}{ 2 \int_{V} \varepsilon_0  |\vec{{\cal E}}|^2 dV}}\sim 0.01-0.02$ \cite{supp_burnett2016} which accounts for the fact that the TLS host material volume $V_h$ may only partially fill the resonator mode volume $V$. We note that the uncertainty in accurately determining $F$ gives a large range for the possible dipole moment ratio $d_F/d_0$. The ranges given for the quantities in the discussion on $d_F$ in the main manuscript are the values obtained for the estimated range of the filling factor.
Notice that in this limit the noise spectrum scales with temperature as ${\propto T^{-(1+2 \mu)}}$. Eq. (\ref{nois}) gives the $1/f$ noise spectrum 
\begin{equation}
\frac{S_{\delta \nu}}{\nu_0^2}=\frac{A_0}{2\pi f}.
\end{equation}

The amplitude $A_0$ can be expressed through the total number of  thermally activated fluctuators ${{\cal N}^{\text{tot}}_F(T)}$ as
\begin{equation}
A_0 \approx \pi \frac{d_F}{d_0} {\cal N}^{\text{tot}}_F(T)
 \left [\chi F^2 P_\gamma \right ] \left (\frac{U_0}{\Gamma_2V_h} \right )^2 
 \label{amp}.
 \end{equation}

\subsection{Number of photons}
\begin{figure}[!t]
	\includegraphics[scale=0.7]{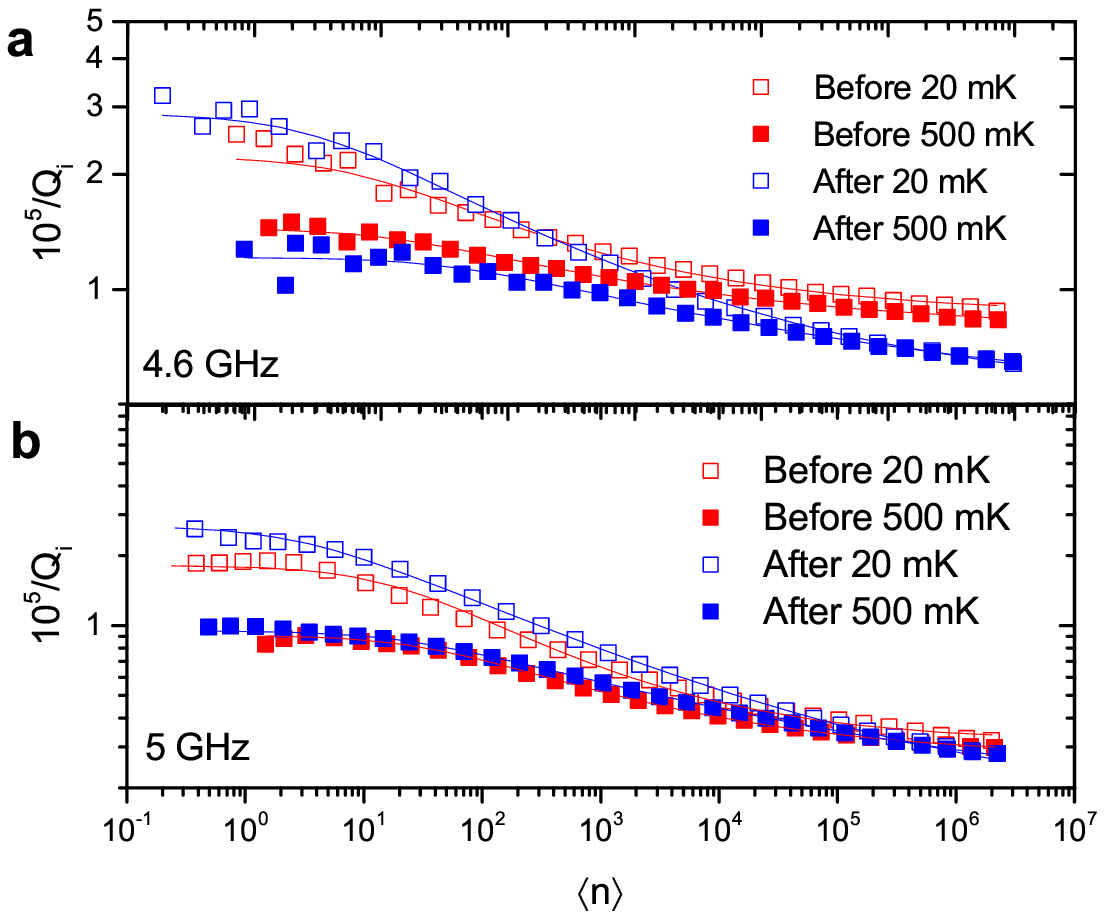}	 	
\vspace{-3mm}	\caption{Inverse internal quality factor as a function of number of photons in the resonator for two extreme temperatures covering those used in all other measurements. Each data point is an increment in applied power by 2 dB, starting at the same low applied power. Fits are to equation (\ref{stmqi}). Some of the data is the same as in Figure \ref{fig:qilog}.} \label{fig:qi}
\end{figure}

Figure \ref{fig:qi}a shows the number of photons in the 4.6 GHz resonator vs internal loss ($Q_i^{-1}$) for the two measurements at two temperatures. Each measurement was made in the same sample cell using the same microwave setup, and the initial assumption is that in the two separate measurements the attenuation in the cryostat was the same. Each data point corresponds to a 2 dB increment in the applied power, both datasets starting at the same low applied power. Therefore, the range of microwave powers applied to the sample is expected to be the same across both measurements. This is further validated by the measurement of white noise levels that are the same (within a factor 2). The white noise level in these measurements is dominated by the microwave power incident on the cryogenic amplifier.

The number of photons within the resonator scales with the loaded quality factor, and therefore also with the internal quality factor. As discussed in the main text, the spin desorption leads to an increase in $Q_i$, meaning that for the same applied microwave power, the number of photons in the resonator is different between the "before" and "after" measurements. Importantly, the noise  scales with both the number of photons within the resonator and with $Q_i$. As is consistent with the literature, we calibrate the applied power such that we compare noise for the same number of photons within the resonator.

\subsection{Power dependence of $Q_i$}
For consistency we here also provide the analysis of the quality factor data within the framework of the STM. Here we expect at strong fields $\langle n\rangle\gg n_c$ 
\begin{equation}
Q_i^{-1} = \frac{F \tan \delta_i }{(1+\langle n\rangle/n_c)^\alpha} + Q_{i,0}^{-1},\label{stmqi}
\end{equation}
where the constant $Q_{i,0}$ accounts for power-independent loss and $n_c$ is a critical photon number for saturation, ${\alpha=0.5}$ and $F$ is the filling factor of the TLS hosting medium in the resonator. By fitting the measured $Q_i(\langle n\rangle)$ to this power law we find $\alpha \sim 0.2$ both before and after spin desorption. Typical fits can be seen in Figure \ref{fig:qi}. 

\subsection{ESR-spectrum and spin density}
The ESR-spectrum in Figure \ref{fig2} is obtained by measuring the transmitted microwave signal, $S_{21}$, around resonance as a function of applied magnetic field using a vector network analyser. 
All noise measurements were performed first, making sure the resonator was not poisoned by vortices. Once noise measurements were completed, the superconducting magnet leads were connected and a magnetic field applied in the plane of the superconducting film. The measured microwave transmission was fitted to\cite{supp_khalil2013}
\begin{equation}
S_{21} = 1-\frac{(1-S_{21, {\rm min}})e^{i\varphi}}{1+2iQ\frac{\nu_0-f}{\nu_0}},\label{eqs3}
\end{equation}
to extract the internal quality factor $Q_i = Q/S_{21, {\rm min}}$. The parameter $\varphi$ accounts for the asymmetry in the resonance line-shape accounting for possible impedance mismatch.  The spin-induced loss is then calculated as $Q_b^{-1}(B)= Q_i^{-1}(B)-Q_i^{-1}(B=0)$. 
We fit the ESR-spectrum to a model of two coupled oscillators to extract the collective coupling, $\Omega$, and line width $\gamma_2 $ ($=1/T_2$ for a Lorentzian ESR peak) of the spin system. 
\begin{equation}
S_{21}(\omega) = 1+\frac{\kappa_c}{i(\omega-\omega_0)-\kappa+\frac{\Omega^2}{i(\omega-\omega_s)-\gamma_2/2}},\label{eqs4}
\end{equation}
Eq. \ref{eqs4} here describes the central $g=2$ peak only and $\omega_0=2\pi\nu_0$ and $\omega_s = g\mu_BB/\hbar$ is the angular resonance frequency and induced Zeeman splitting of the spins respectively, and $\kappa_{(c)} = \omega_0/Q_{(c)}$.
From the collective coupling $\Omega$ we can evaluate the surface spin density based on the geometry of the resonator\cite{supp_degraaf2017}. Comparing the same resonator before and after annealing also gives a direct measure of the relative reduction in spin density independent of resonator geometry via the observed reduction in collective coupling of the spins, $\Omega\propto \sqrt{n}$. In the 'After' measurement we have removed $\sim 2\cdot 10^{17}$ Hydrogen spins$/$m$^2$ and the density of $g=2$ spins was reduced 5.3 times to $\sim 0.17\cdot 10^{17}$ spins$/$m$^2$. Figure \ref{fig2}e shows the good agreement of the ESR data to theory. 
We note that the reduction of 5.3 times is larger than previously observed\cite{supp_degraaf2017}, suggesting that the $g=2$ spins have a larger desorption energy than the hydrogen. 

\subsection{CW power saturation measurements: $T_1$}
\begin{figure}[!b]
	\includegraphics[scale=0.66]{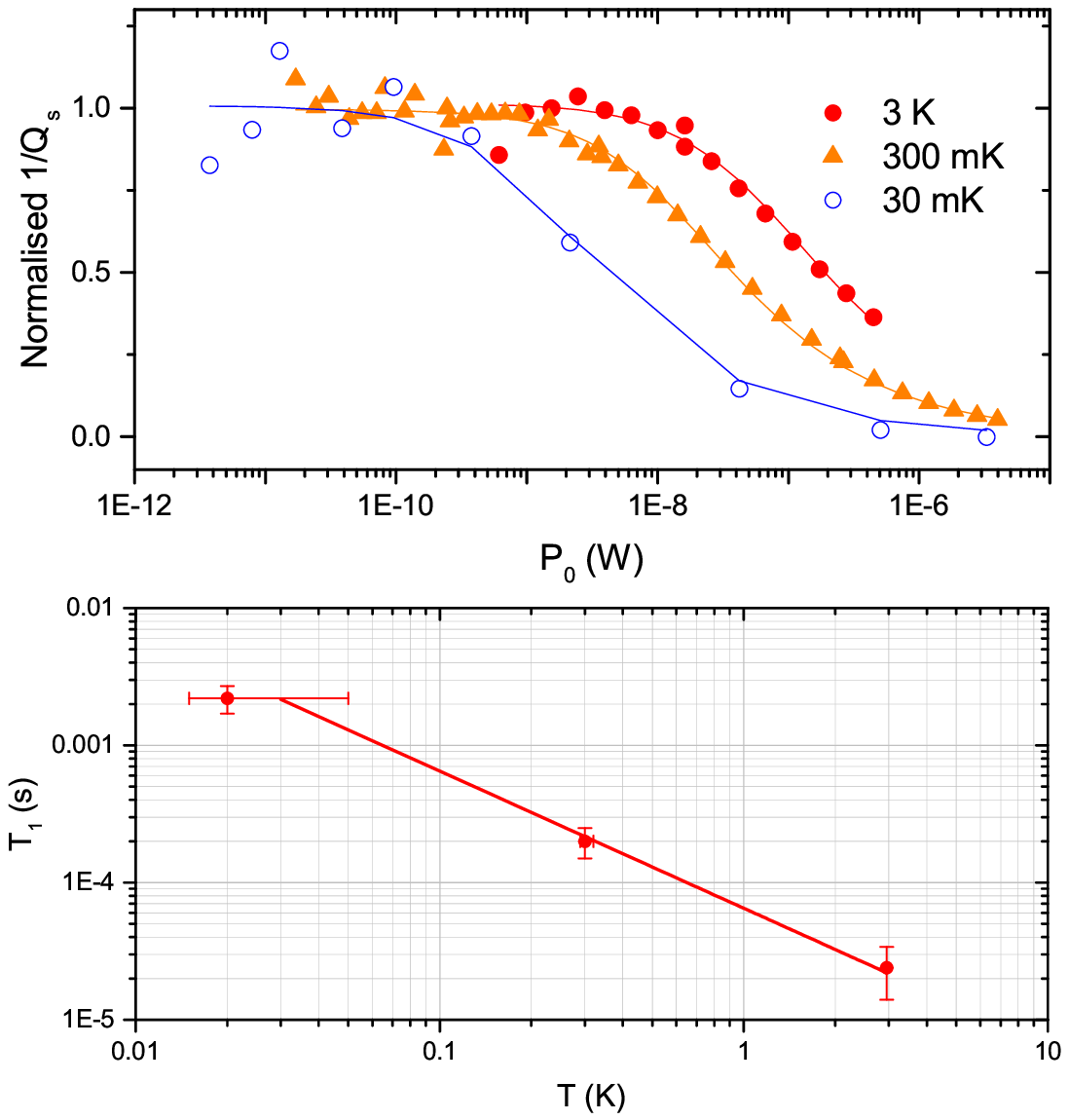}	 	
\vspace{-3mm}	\caption{Spin relaxation times. a) CW power saturation measurements at three different temperatures showing the normalised inverse dissipation into the spin system as a function of circulating power in the 4.6 GHz resonator after annealing. b) the extracted (from fits) relaxation time $T_1$ as a function of temperature for the $g=2$ peak. Solid line is $1/T$.} \label{fig:t1}
\end{figure}

When evaluating the spin density it is essential to ensure that the spin ensemble is not saturated by the microwave signal in the resonator. To verify this we measure the ESR-spectrum at a wide set of applied powers and extract the dissipation into the spin system at the $g=2$ peak as a function of circulating power in the resonator. The result for one such measurement (after annealing, evaluated for the $g=2$ peak) is shown in Figure \ref{fig:t1}a for three different temperatures. 
The method to evaluate $Q_s$ is described in Ref. \cite{supp_degraaf2017} together with the methodology used to extract the spin relaxation time $T_1$ plotted versus temperature in Figure \ref{fig:t1}b. Interestingly we find a $T^{-1}$ dependence of the relaxation time, a signature of direct spin-lattice relaxation as the dominant mechanism for spin energy relaxation\cite{supp_schweiger}. Direct phonon relaxation and a $T_1\propto T^{-1}$ dependence is also the dominant mechanism for TLS relaxation in amorphous glasses at low temperatures, well captured by both the STM and the GTM\cite{supp_faoro2015}, predicting a similar dominating phonon relaxation time in the ms range.

We note that spin desorption does not change $T_1$, while the electron spin dephasing time $T_2$ inferred from the transition line-width increases marginally (table \ref{tab1}), an indication of reduced spin-spin induced decoherence, alternatively the remaining spins could be of a different nature.

\subsection{Noise measurement setup: Dual Pound locking}
The measurement setup we use is a further development of the Pound locking technique\cite{supp_tobiaspound} for microwave resonators. This modification allows us to simultaneously measure the frequency noise in two different resonators, increasing the amount of data collected and allowing for measurement of correlated noise. 

\begin{figure}
	\includegraphics[scale=0.46]{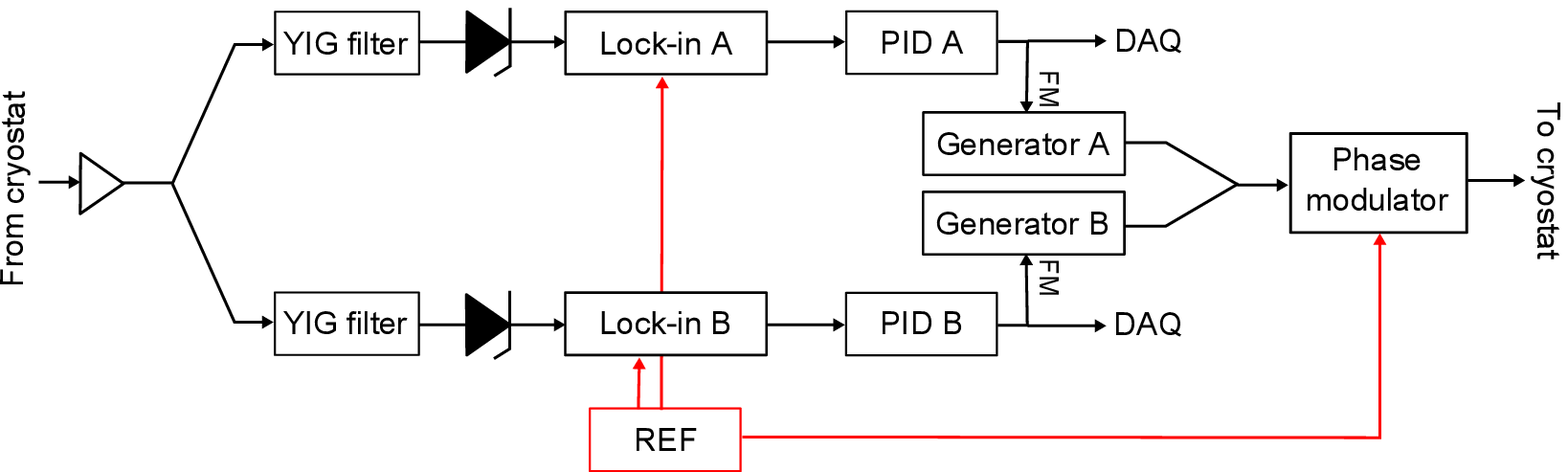}	 	
	\caption{Dual Pound-locking measurement setup. For details see text.} \label{fig:setup}
\end{figure}

Pound locking is a highly accurate technique to directly measure frequency noise of microwave oscillators. 
This as opposed to measuring the phase noise $S_\varphi$ using a homo/heterodyne technique\cite{supp_barendsnoisepaper}. The advantage is that we gain in sensitivity and the measurement does not suffer from additional complications such as the Leeson effect, and it is especially useful in cryogenic environments, where homo- or heterodyne techniques suffer from a wide range of fluctuations, such as in electrical length in each of the two measurement paths (signal and reference), and thermal fluctuations. The Pound locking technique instead sends the signal and reference through the same physical transmission line, where the reference takes the form of a phase modulated spectrum on top of the signal, making the measurement insensitive to first order in any variations in electrical length. The phase modulation frequency is recovered by a non-linear detector (here a diode) and any deviations in the signal frequency from the resonator frequency causes a beating at the phase modulation frequency. This beating is nulled using a lock-in in series with a PID controller which adjusts the signal frequency sent out by the microwave generator to match the instant resonance frequency.

Instead of a single Pound loop we here run two loops in parallel, as shown in Figure \ref{fig:setup}. Each loop, A and  B, works in the same way as described in detail in Ref. \cite{supp_tobiaspound}, locked to the 4.6 and 5.0 GHz resonator respectively. The microwave signals from each loop are combined and sent through the same transmission line in the cryostat, and later selectively split to each arm using 7th order tunable YIG filters with a bandwidth of $40$ MHz tuned to each respective resonance frequency. This type of multiplexed setup in principle allows for an arbitrary number of Pound loops in parallel without introducing any cross-coupling and errors in frequency measurement, as long as the YIG filters can selectively isolate the phase modulation spectrum from each measured resonator.

The power applied in each Pound loop was carefully verified using a spectrum analyser and adjusted to be equal in the two measurements, both for power incident on the resonators and power incident on the detector diode.

\subsection{Loss tangent measurements}
\begin{figure}
	\includegraphics[scale=0.32]{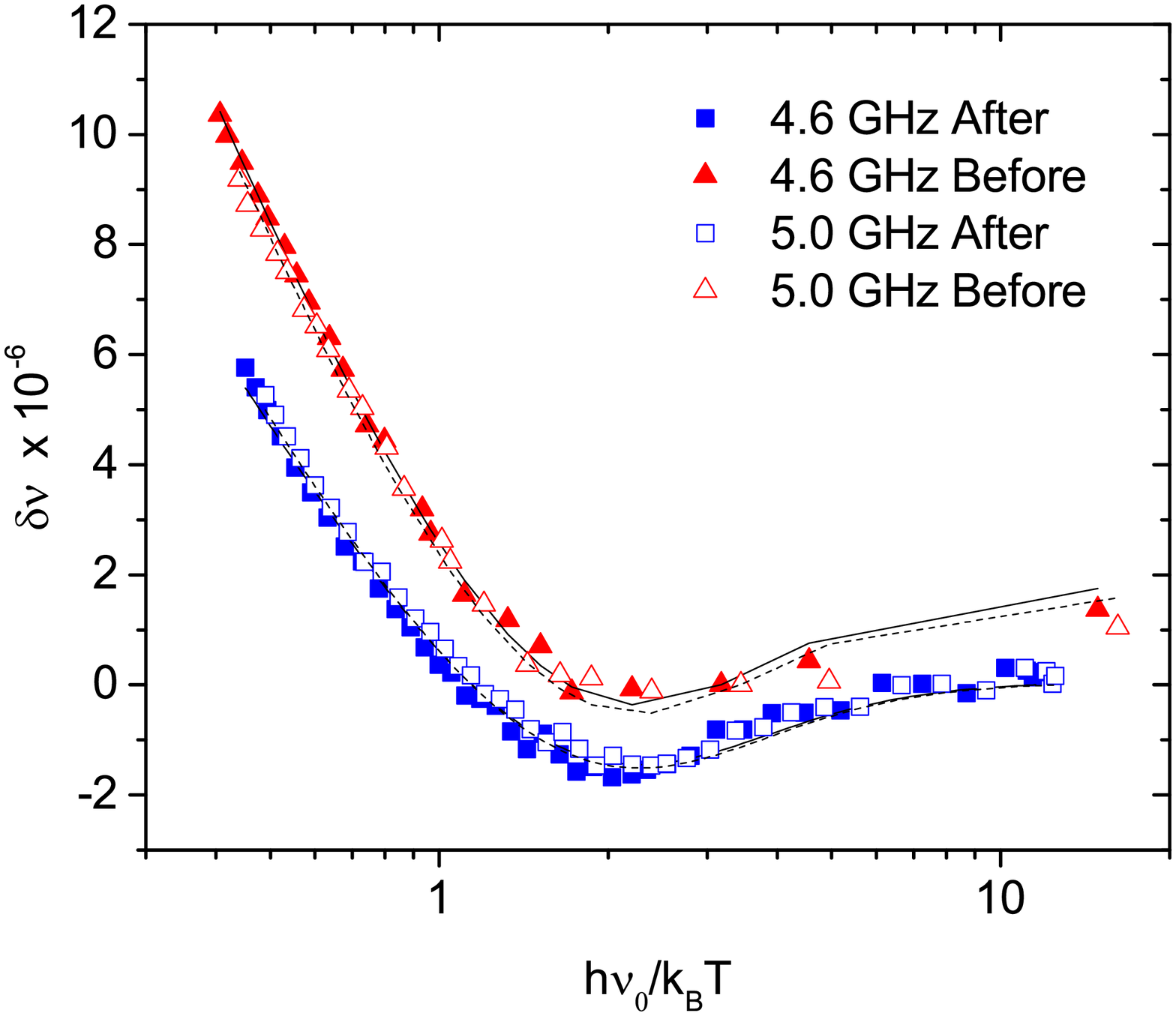}	 	
	\caption{Intrinsic loss tangent. Frequency shift of the resonators as a function of temperature and fits to Eq. (\ref{eq:tand}) (black lines). For both resonators we observe a reduction in loss tangent upon surface spin desorbtion by 25-30\%. Extracted values for $\tan\delta_i$ are shown in Table 1. Curves are offset for clarity.} \label{fig:tand}
\end{figure}

To obtain the loss tangent we measure the frequency shift of each resonator while slowly ramping up the temperature of the cryostat over the course of $\sim 120$ minutes. The frequency is measured using the Pound-loop. The loss tangent $\tan\delta_i$ is then extracted from fits of the $\nu_0(T)$ data to the STM (and GTM).
\begin{eqnarray}
\delta \nu(T) &=& F \tan\delta\bigg[{\rm Re}\bigg(\Psi(\frac{1}{2}+\frac{\nu_0h}{2\pi i\nonumber k_BT})\\&&+\Psi(\frac{1}{2}+\frac{\nu_0h}{2\pi i k_BT_0})-\ln\frac{T}{T_0}\bigg) \bigg].\label{eq:tand}
\end{eqnarray}
Here $\delta \nu(T) = (\nu_0-\nu(T))/\nu_0$, $\Psi$ is the di-gamma function and $T_0$ is a reference temperature. The measured data and fits to Eq. (\ref{eq:tand}) are shown in Figure \ref{fig:tand}. Extracted parameters are shown in Table \ref{tab1}.

\subsection{Noise analysis}
\begin{figure}
	\hspace{-10mm}\includegraphics[scale=1.7]{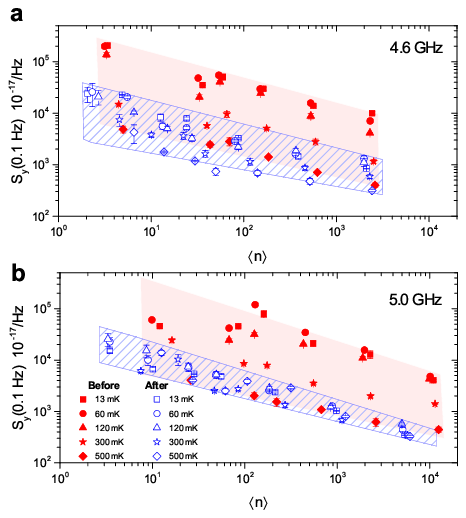}	 	
	\caption{Reduction of noise due to spin desorption.  Frequency noise power spectral density $S_y(f) = S_{\delta \nu}(f)/\nu_0^2$ at $f=0.1$ Hz for the a) $\nu_0=4.6$ GHz resonator (same data as in Figure \ref{fig2}a) and b) the  5.0 GHz resonator.  Red solid markers are before, and blue hollow markers are after spin desorption respectively. Shaded regions are a guide for the eye.}
	\label{fig2supp}
\end{figure}
The sampled frequency vs time signal recorded from the Pound loop is converted to frequency noise spectral density $S_y$ by calculating the overlapping Allan-variance $\sigma_y^2(\tau)$ (AVAR) for M discrete samplings $f_k(n\tau)$ at multiples $n$ of the sampling rate $\tau$. 
\begin{equation}
\sigma_y^2(n\tau) = \frac{1}{2(M-1)}\sum_{k=1}^{M-1}(f_{k+1}-f_k)^2
\end{equation}
For $1/f$ noise the  spectral density $S_y(f) = h_{-1}/f$ relates to the Allan variance $\sigma_y^2 = 2\ln{(2)}h_{-1}$ via the coefficient $h_{-1}$ \cite{rubiola}.
The AVAR is evaluated at several time-scales $t=n\tau$ ranging from 20 to 80 seconds, well within the 1/f noise limit, and the average value for $h_{-1}$ is obtained with high statistical significance. Error bars are calculated from the standard deviation of the multiple evaluations of the AVAR in the same time interval. Each datapoint in Figure \ref{fig2} is the result of a 2.8 hours long measurement, collecting $10^5$ samples without interruption at a rate $\tau^{-1}=10$ Hz. Such long measurement times are required to obtain statistically significant results for $h_{-1}$ since the $1/f$ noise in these high-Q resonators is only exceeding the system white noise level at frequencies below $\sim 1-0.1$ Hz, in particular at high temperatures and low applied powers and especially in the 'After' measurement where the $1/f$ noise level is significantly lower. 

Full data for both resonators measured is shown in Figure \ref{fig2supp}.

\subsection{Temperature dependence of $S_y$}
\begin{figure}[!tb]
	\includegraphics[scale=0.34]{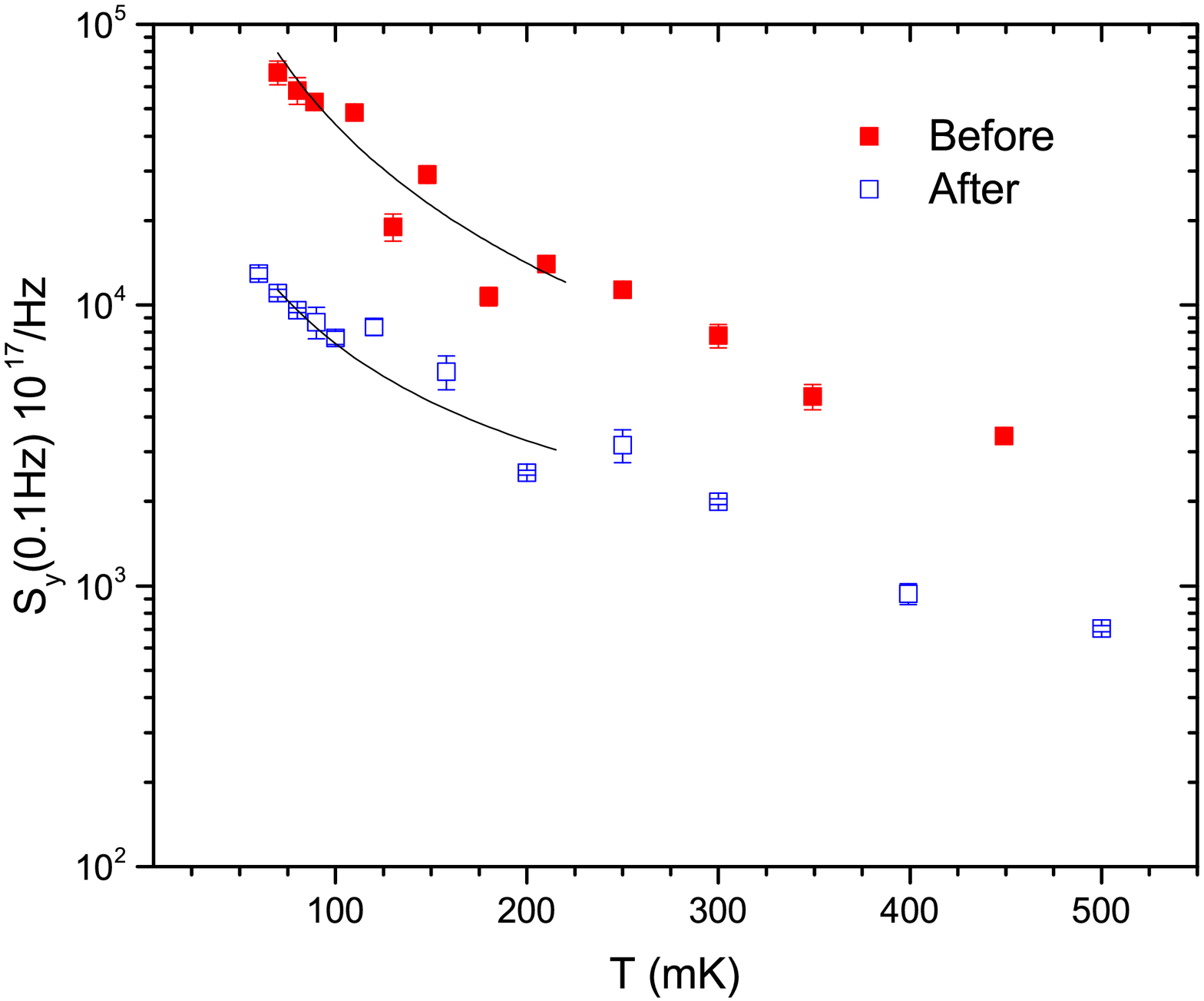}	 	
	\caption{Temperature dependence of the noise power spectral density at $\langle N\rangle = 8\pm 2$ before and $\langle N\rangle = 4\pm 2$ after surface spin desorption. Solid lines are fits to $S_y(T) = A_0T^{-(1+2 \mu)}$. Before annealing we find $2 \mu = 0.64 \pm 0.50$ while after annealing $2 \mu = 0.43\pm0.21$.  Due to thermal saturation, data points below 70 mK are excluded from the fit. We also only consider the low temperature regime $k_BT<\hbar\omega_0\approx 220 $ mK.} \label{fig:mu}
\end{figure}

To extract $\mu$ we measure the temperature dependence of $S_y$, which in the low power and low temperature limit is expected to scale as $S_y(T) \propto A_\mu T^{-(1+2\mu)}$. The low temperature limit is given by $T < h\nu/k_B\approx 220 $ mK for our $\nu = 4.6$ GHz resonator. This measurement and fits to extract $\mu$ are shown in Figure \ref{fig:mu}. Confidence intervals given for $\mu$ are propagated error bars from the calculation of $S_y$. Indeed we find $\mu>0$ for both measurements and whilst error bars are relatively large, we conclude that interaction is still present and $\mu$ has not changed by a significant amount.  We do not have data at low enough photon numbers to accurately evaluate $\mu$ for the 5 GHz resonator.

\subsection{Correlated noise}
\begin{figure}[!t]
	\includegraphics[scale=0.34]{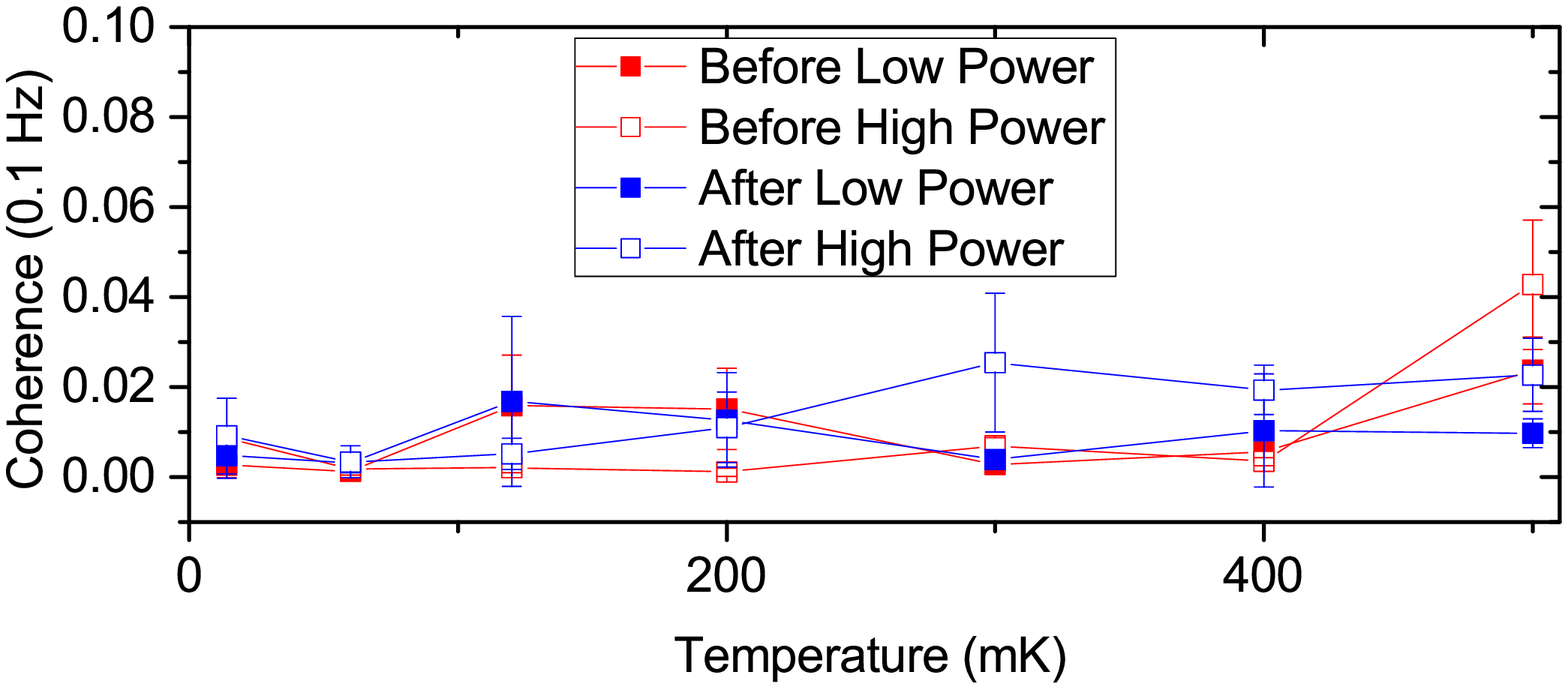}	 	
	\caption{Uncorrelated noise. The coherence (normalised correlation) at 0.1 Hz between simultaneously measured 4.6 and 5.0 GHz resonators. The measurement shows that the 1/f noise in each resonator at the time-scale of 0.1 Hz is dominated by local sources at all relevant temperatures. Low power is equivalent to $\langle N\rangle\sim 1-10$ and high power corresponds to $\langle N\rangle\sim 10^3-10^4$.} \label{correlation}
\end{figure}

To rule out external factors, such as system noise, magnetic field or thermal fluctuations, vibrations, and vortices influencing the results we verify that the measured $1/f$ noise is local to each resonator by measuring their correlated noise. We evaluate the correlated noise as the coherence function from the spectral densities
\begin{equation}
C = \frac{|S_{AB}|^2}{S_{AA}S_{BB}}.
\end{equation}
Here $S_{XY}$ is the cross-power spectral density 
\begin{equation}
S_{XY} = \int_{-\infty}^\infty dt e^{-i\omega t}\int_{-\infty}^\infty d\tau \nu_X(\tau) \nu_Y(t+\tau)
\end{equation}
of frequency fluctuations $\nu_A(t)$ and $\nu_B(t)$, where A and B denote the two different resonators.
Figure \ref{correlation} shows the measured coherence $C(0.1 {\rm Hz})$ as a function of temperature and for the two extreme powers applied to each resonator, obtained from the same data as in Figure \ref{fig2supp}. We observe no correlations at the time-scale of 0.1 Hz that is relevant for the $1/f$ noise analysis performed in this work.

\begin{figure}
	\includegraphics[scale=0.68]{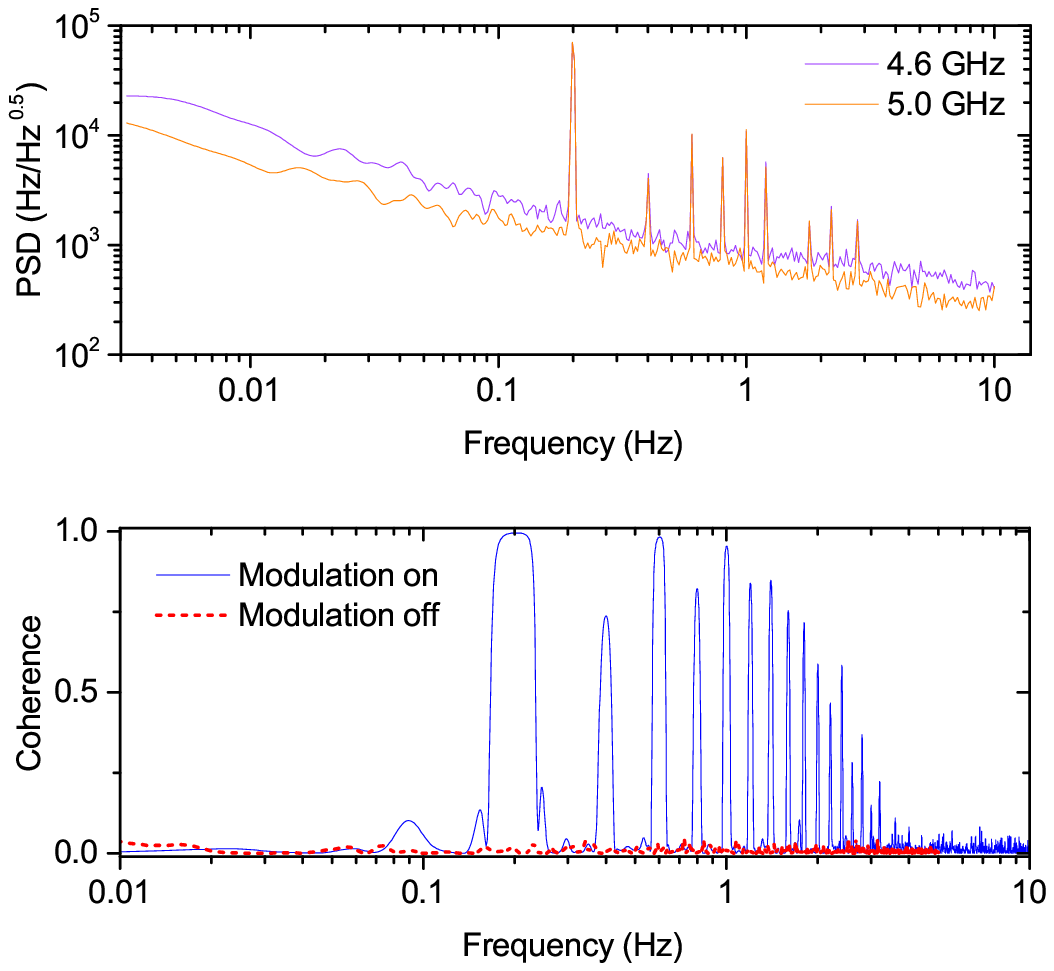}	 	
	\caption{Correlated noise due to external perturbation. a) Power spectral density of frequency fluctuations in the two resonators (before spin desorption, at $T=20$ mK) when a small oscillating magnetic field at 0.2 Hz is applied to the sample. The coherence data for the same measurement as in a) and another measurement when the field modulation is turned off. The amplitude of field fluctuations was $0.1$ Oe resulting in approximately 20 kHz oscillations in the resonance frequencies of the resonators.} \label{correlationverification}
\end{figure}

As another control experiment we measured the coherence while applying a weak (0.02 mT) external magnetic field perpendicular to the superconducting thin-film plane of the sample at a frequency of 0.2 Hz. The measured coherence is very strong at this particular frequency and its higher harmonics, as shown in Figure \ref{correlationverification}. These measurements clearly verify that we have successfully eliminated any common sources of noise and the dominating contribution to the $1/f$ noise originates from noise sources local to each resonator within the entire measurement space presented in this work.

{\footnotesize
	\begin{thebibliography}{999}
		
	\bibitem{devoret2013} Devoret, M. H. and Schoelkopf, R. J. Superconducting circuits for quantum information: an outlook. \href{http://dx.doi.org/10.1126/science.1231930}{Science {\bf 339}, 1169 (2013)}.
	
		\bibitem{paladino2014} Paladino, E.,  Galperin, Y. M., Falci, G. and Altshhuler, B. L.  1/f noise: Implications for solid-state quantum information.
		\href{http://dx.doi.org/10.1103/RevModPhys.86.361}{Rev. Mod. Phys. {\bf 86}, 361 (2014)}.
		
		\bibitem{muller2017} M\"uller, C., Cole, J. H. and Lisenfeld, J. Towards understanding two-level-systems in amorphous solids - insights from quantum devices. \href{http://arxiv.org/abs/1705.01108}{arXiv preprint 1705.01108v1 (2017)}.
		

		
		
	\bibitem{anderson} Anderson, P. Halperin, B. and Varma, C. Anomalous low-temperature thermal properties of glasses and spin glasses. \href{http://dx.doi.org/10.1080/14786437208229210}{Philos. Mag. {\bf 25}, 1 (1972)}.
	
	\bibitem{phillips} Phillips, W. A.  Two-level states in glasses. \href{http://dx.doi.org/10.1088/0034-4885/50/12/003}{Rep. Prog. Phys. {\bf 50}, 1657 (1987)}.
		
		
		
	\bibitem{pearl2016} 
		Tenorio-Pearl, J. O., Herbschleb, E. D., Fleming, S., Creatore, C., Oda, S, Milne, W. I. and Chin, A. W. Observation and coherent control of interface-induced electronic resonances in a field-effect transistor.  \href{http://dx.doi.org/10.1038/nmat4754}{ Nature Materials {\bf 16}, 208 (2016)}.
		
		\bibitem{arsalan2014} Pourkabirian, A. Gustafsson, M. V. Johansson, G.,  Clarke, J. and Delsing, P.  
		Nonequilibrium probing of two-level charge fluctuators using the step response
		of a single-electron transistor.  \href{http://dx.doi.org/10.1103/PhysRevLett.113.256801}{ Phys. Rev. Lett. {\bf 113}, 256801 (2014)}.
		
		\bibitem{iontrap} Brownnutt, M., Kumph, M., Rabl, P. and Blatt, P. Ion-trap measurements of electric-field noise near surfaces. \href{http://dx.doi.org/10.1103/RevModPhys.87.1419}{Rev. Mod. Phys. {\bf 87}, 1419 (2015).}
		
		\bibitem{kumar2016} Kumar, P. et al.,  Origin and suppression of 1/f magnetic flux noise.  
		\href{https://doi.org/10.1103/PhysRevApplied.6.041001}{Phys. Rev. Appl. {\bf 6}, 041001 (2016)}.
		
		\bibitem{quintana2017} Quintana, C. M. et al. Observation of classical-quantum crossover of 1/f flux noise	and its paramagnetic temperature dependence. 
		\href{https://doi.org/10.1103/PhysRevLett.118.057702}{Phys. Rev. Lett. {\bf 118}, 057702 (2017)}.
			
			
			
			
	\bibitem{steffen2017}Steffen, M., Sandberg, M. and Srinivasan, S. 
	Recent research trends for high coherence quantum circuits.  \href{http://dx.doi.org/10.1088/1361-6668/aa55a2}{Supercond. Sci. Technol.  {\bf 30}, 030301 (2017)}.
		
		
		
		
	\bibitem{chu2016} Chu, Y., Axline, C., Wang, C., Brecht, T., Gao, Y. Y., Frunzio, L. and Schoelkopf, R. J. Suspending superconducting qubits by silicon micromachining. \href{http://dx.doi.org/10.1063/1.4962327}{Appl. Phys. Lett. {\bf 109}, 112601 (2016)}.
		
	\bibitem{burnett2014} Burnett, J. et al.,  Evidence for interacting two-level systems from the 1/f noise of a superconducting resonator. 
	\href{http://dx.doi.org/10.1038/ncomms5119}{Nature Communications {\bf 5}, 4119 (2014)}.
		
		

		
	\bibitem{neill2013} Neill, C. et al. Fluctuations from edge defects in superconducting resonators. \href{http://dx.doi.org/10.1063/1.4818710}{Appl. Phys. Lett. {\bf 103}, 072601 (2013)}.
	
	
	\bibitem{burin2015} Burin, A. L., Matityahu, S. and Schechter, M. Low-temperature 1/f noise in microwave dielectric constant of amorphous dielectrics in Josephson qubits. \href{http://dx.doi.org/10.1103/PhysRevB.92.174201}{Phys. Rev. B {\bf 92}, 174201 (2015)}.
			
	\bibitem{faoro2015} Faoro, L. and Ioffe, L. B. Interacting tunneling model for two-level systems in amorphous materials and its predictions for their dephasing and noise in superconducting microresonators.
	\href{http://dx.doi.org/10.1103/PhysRevB.91.014201}{Phys. Rev. B {\bf 91}, 014201 (2015)}.
			
	\bibitem{skacel2014} Skacel, S. T. et al. Probing the TLS density of states in SiO films using superconducting lumped element resonators. \href{http://dx.doi.org/10.1063/1.4905149}{Appl. Phys. Lett. {\bf 106}, 022603 (2015)}.
				
	\bibitem{burnett2016} Burnett, J., Faoro, L. and Lindstr\"om, T.  Analysis of high quality superconducting resonators: consequaneces for TLS properties in amorphous oxides. 
	\href{http://dx.doi.org/10.1088/0953-2048/29/4/044008}{Supercond. Sci. Technol. {\bf 29}, 044008 (2016)}.
				
			
	\bibitem{lisenfeld2010} Lisenfeld, J., M\"uller, C., Cole, J. H., Bushev, P., Lukashenko, A., Shnirman, A. and  Ustinov, A. V. Measuring the temperature dependence of individual two-level systems by direct coherent control. \href{http://dx.doi.org/10.1103/PhysRevLett.105.230504}{ Phys. Rev. Lett. {\bf 105}, 230504 (2010)}.
		

	
	

	\bibitem{khalil2010} 
	Khalil, M. S., Wellstood, F. C. and Osborn, K. D. Loss Dependence on Geometry and Applied
	Power in Superconducting Coplanar Resonators. \href{http://dx.doi.org/}{ IEEE Trans. Appl. Sup. {\bf 21}, 879 (2011)}.
		
	
	\bibitem{lindstrom2009} Lindstr\"om, T., Healey, J., Colclough, M., Muirhead, C. and Tzalenchuk, A. Y.
	Properties of superconducting planar resonators at millikelvin temperatures. \href{https://doi.org/10.1103/PhysRevB.80.132501}{Phys. Rev. B {\bf 80}, 132501 (2009)}.
		
		
	\bibitem{faoro2012} Faoro, L. and  Ioffe, L. B. Internal loss of superconducting resonators induced by interacting two-level systems. 
	\href{http://dx.doi.org/10.1103/PhysRevLett.109.157005}{Phys. Rev. Lett. {\bf 109}, 157005 (2012)}.
		
	
	\bibitem{lisenfeld2015} Lisenfeld, J., Grabovskij, G. J.,  M\"uller, C.,  Cole, J. H., Weiss, G. and Ustinov, A. V. Observation of directly interacting coherent two-level systems in an amorphous material. \href{https://doi.org/10.1038/ncomms7182}{Nature Communications {\bf 6}, 6182 (2015)}. 
	
	\bibitem{holder2013} Holder, A. M. Osborn, K. D.  Lobb, C. J. and Musgrave, C. B.  Bulk and surface tunneling hydrogen defects in alumina. 
	\href{http://dx.doi.org/10.1103/PhysRevLett.111.065901}{Phys. Rev. Lett. {\bf 111}, 065901 (2013)}.
		
	\bibitem{lee2014} Lee, D., DuBois, J. L.  and Lordi, V. Identification of the local sources of paramagnetic noise in superconducting qubit devices fabricated on $\alpha$-Al$_2$O$_3$ substrates using density-functional calculations. 
	\href{http://dx.doi.org/10.1103/PhysRevLett.112.017001}{Phys. Rev. Lett, {\bf 112}, 017001 (2014)}.
		
	\bibitem{wang2015} Wang, H., Shi, C.,  Hu, J., Han, S., Yu, C. C. and Wu, R. Q. Candidate source of flux noise in SQUIDs: Adsorbed oxygen molecules. 	
	\href{http://dx.doi.org/10.1103/PhysRevLett.115.077002}{Phys. Rev. Lett. {\bf 115}, 077002 (2015)}.
		
	\bibitem{lordi2017} Adelstein, N., Lee., D., DuBois, J. L., Varley, J. B. and Lordi, V. magnetic stability of oxygen defects on the SiO2 surface. \href{http://dx.doi.org/10.1063/1.4977194}{AIP Advances, {\bf 7}, 025110 (2017)}.
		
		
	\bibitem{desousa2007}  de Sousa, R. Dangling-bond spin relaxation and magnetic 1/f noise from the amorphous-semiconductor/oxide interface: theory.
	\href{http://dx.doi.org/10.1103/PhysRevB.76.245306}{Phys Rev. B {\bf 76}, 245306 (2007)}.
		

		
	\bibitem{degraaf2017} de Graaf, S. E.,  Adamyan, A. A., Lindstr\"om, T., Erts, D., Kubatkin, S. E., Tzalenchuk,  A. Ya. and Danilov, A. V. Direct identification of dilute surface spins on Al2O3: origin of flux noise in quantum circuits. 
	\href{http://dx.doi.org/}{Phys. Rev. Lett. {\bf 118}, 057703 (2017)}.
		
	\bibitem{kelber2007} Kelber, J. A. Alumina surfaces and interfaces under non-ultrahigh vacuum conditions. \href{http://dx.doi.org/10.1016/j.surfrep.2006.12.003}{Surf. Sci. Rep. {\bf 62}, 271-303 (2007)}.
		
	\bibitem{sendelbach2008} Sendelbach, S., Hover, D.,  Kittel, A., M$\rm\ddot{u}$ck, M.,  Martinis, J. M.  and  McDermott, R.  Magnetism in SQUIDs at milliKelvin temperatures. 
	\href{http://dx.doi.org/10.1103/PhysRevLett.100.227006}{ Phys. Rev. Lett., {\bf 100}, 227006 (2008)}.
		
	

\bibitem{samkharadze2016} Samkharadze, N.,  Bruno, A., Scarlino, P., Zheng, G.,  DiVincenzo, D. P.,  DiCarlo, L. and  Vandersypen, L. M. K.
High-kinetic-inductance superconducting nanowire resonators for circuit QED in a magnetic field. \href{https://doi.org/10.1103/PhysRevApplied.5.044004}{Phys. Rev. Applied {\bf 5}, 044004 (2016)}. 

		
	\bibitem{aplpaper} de Graaf, S. E.,  Davidovikj, D., Adamyan, A., Kubatkin, S. E. and  Danilov, A. V. Galvanically split superconducting microwave resonators for introducing internal voltage bias. 
	\href{http://dx.doi.org/10.1063/1.4863681}{Appl. Phys. Lett. {\bf 104}, 052601 (2014)}.

	\bibitem{jap2012} de Graaf, S.E., Danilov, A.V., Adamyan, A., Bauch, T., Kubatkin, S.E. Magnetic field resilient superconducting fractal resonators for coupling to free spins, \href{http://dx.doi.org/10.1063/1.4769208}{J. Appl. Phys., {\bf 112}, 123905 (2012)}.
	
	\bibitem{tobiaspound} Lindstr\"om, T., Burnett, J., Oxborrow, M. and  Tzalenchuk, A. Ya. Pound-locking for characterization of superconducting microresonators. 
	\href{http://dx.doi.org/10.1063/1.3648134}{Rev. Sci. Instrum. {\bf 82}, 104706 (2011)}.
		
					
		
	\bibitem{hass1998} Hass, K. C., Schneider, W. F.,  Curioni, A. and  Andreoni, W. The chemistry of water on alumina surfaces: Reaction dynamics from first principles.		\href{http://dx.doi.org/10.1126/science.282.5387.265}{Science {\bf 282}, 265-268 (1998)}. 
			
	\bibitem{lisenfeld2016} Lisenfeld, J. et al. Decoherence spectroscopy with individual two-level tunneling defects. \href{http://dx.doi.org/10.1038/srep23786}{Scientific Reports {\bf 6}, 23786 (2016)}.
	
	
	\bibitem{martinis2004} Martinis, J. M. et al. decoherence in Josephson qubits from dielectric loss. \href{http://dx.doi.org/10.1103/PhysRevLett.95.210503}{Phys. Rev. Lett. {\bf 95}, 210503 (2005)}.
			
	\bibitem{sarabi2016} Sarabi, B.,  Ramanayaka, A. N., Burin,  A. L.,  Wellstood, F. C. and  Osborn, K. D. Projected dipole moments of individual two-level defects extracted using circuit quantum electrodynamics. \href{http://dx.doi.org/10.1103/PhysRevLett.116.167002}{Phys. Rev. Lett. {\bf 116}, 167002 (2016)}.
		
	\bibitem{bedilo2014}  Bedilo, A. F.,  Shuvarakova,   E. I.,  Rybinskaya, A. A., and Medvedev, D. A. Characterization of electron donor and electron acceptor sites on the surface of sulfated alumina using spin probes.  
	\href{http://dx.doi.org/10.1021/jp503523k}{J. Phys. Chem. C {\bf 118}, 15779 (2014)}.
		
	\bibitem{astafiev2004} Astafiev, O.,  Pashkin, Yu. A., Nakamura, Y., Yamamoto, T. and Tsai, J. S. Quantum noise in the Josephson charge qubit.
	\href{http://dx.doi.org/10.1103/PhysRevLett.93.267007}{ Phys. Rev. Lett. {\bf 93}, 267007 (2004)}.



	\end{thebibliography}
}


{\footnotesize
	\begin{thebibliography}{99}

		
		\bibitem{supp_faoro2015} Faoro, L. and Ioffe, L. B. Interacting tunneling model for two-level systems in amorphous materials and its predictions for their dephasing and noise in superconducting microresonators.
		\href{http://dx.doi.org/10.1103/PhysRevB.91.014201}{Phys. Rev. B {\bf 91}, 014201 (2015)}.
		
		
		\bibitem{supp_lisenfeld2016} Lisenfeld, J. et al. Decoherence spectroscopy with individual two-level tunneling defects. \href{http://dx.doi.org/10.1038/srep23786}{Scientific Reports {\bf 6}, 23786 (2016)}.
		
		\bibitem{supp_burnett2016} Burnett, J., Faoro, L. and Lindstr\"om, T.  Analysis of high quality superconducting resonators: consequaneces for TLS properties in amorphous oxides. 
		\href{http://dx.doi.org/10.1088/0953-2048/29/4/044008}{Supercond. Sci. Technol. {\bf 29}, 044008 (2016)}.
		
		\bibitem{supp_khalil2013} Khalil, M. S. et al. Evidence for hydrogen two-level systems in atomic layer deposition oxides. \href{http://dx.doi.org/10.1063/1.4826253}{Appl. Phys. Lett. {\bf 103}, 162601 (2013)}.
			
		\bibitem{supp_degraaf2017} de Graaf, S. E.,  Adamyan, A. A., Lindstr\"om, T., Erts, D., Kubatkin, S. E., Tzalenchuk,  A. Ya. and Danilov, A. V. Direct identification of dilute surface spins on Al2O3: origin of flux noise in quantum circuits. 
		\href{http://dx.doi.org/}{Phys. Rev. Lett. {\bf 118}, 057703 (2017)}.
		
		\bibitem{supp_schweiger} A. Schweiger, G. Jeschke, Principles of pulse electron paramagnetic resonance, Oxford University Press (New York 2001).
		
		
		\bibitem{supp_tobiaspound} Lindstr\"om, T., Burnett, J., Oxborrow, M. and  Tzalenchuk, A. Ya. Pound-locking for characterization of superconducting microresonators. 
		\href{http://dx.doi.org/10.1063/1.3648134}{Rev. Sci. Instrum. {\bf 82}, 104706 (2011)}.
		
	
		
		\bibitem{supp_barendsnoisepaper} Barends, R., Hortensius, H. L., Zijlstra, T., Baselmans, J. J. A., Yates, S. J. C., Gao, R. J. and Klapwijk, T. M. Noise in NbTiN, Al, and Ta superconducting resonators on silicon and sapphire substrates. \href{http://dx.doi.org/10.1109/TASC.2009.2018086}{IEEE Trans. Appl. Supercond. {\bf 19}, 936 (2009)}.
		
		\bibitem{rubiola} Rubiola, E. Phase noise and frequency stability in oscillators  (Cambridge University Press, 2009).



	\end{thebibliography}
}
\end{document}